\documentclass[modern]{aastex631}

\usepackage{hyperref}
\let\tablenum\relax 
\usepackage{siunitx}
\usepackage{graphicx}%
\usepackage{multirow}%
\usepackage{amsmath,amssymb,amsfonts}%
\usepackage{amsthm}%
\usepackage{mathrsfs}%
\usepackage[title]{appendix}%
\usepackage{xcolor}%
\usepackage{textcomp}%
\usepackage{booktabs}%
\usepackage{algorithmicx}%
\usepackage{algpseudocode}%
\usepackage{listings}%

\begin{document}

\title{The Blinking Crystallinity of Europa: A Competition between Irradiation and Thermal Alteration.}

\correspondingauthor{Cyril Mergny}

\author[0009-0002-1910-6991]{Cyril Mergny}
\affiliation{Université Paris-Saclay, CNRS, GEOPS \\
Orsay, 91405, France}
\email{cyril.mergny@universite-paris-saclay.fr}

\author[0000-0002-2857-6621]{Frédéric Schmidt}
\affiliation{Université Paris-Saclay, CNRS, GEOPS \\
Orsay, 91405, France}
\affiliation{Institut Universitaire de France \\
Paris, France}

\author{Felix Keil}
\affiliation{Université Paris-Saclay, CNRS, GEOPS \\
Orsay, 91405, France}

\begin{abstract}
The surface of Europa experiences a competition between thermally-induced crystallization and radiation-induced amorphization processes, leading to changes of its crystalline structure.
The non-linear crystallization and temperature-dependent amorphization rate, incorporating ions, electrons and UV doses, are integrated into our multiphysics surface model (MSM) \textit{LunaIcy}, enabling simulations of these coupled processes on icy moons.

Thirty simulations spanning $\SI{100 000}{years}$, covering the full ranges of albedo and latitude values on Europa, explore the competition between crystallization and irradiation. 
This is the first modeling of depth-dependent crystallinity profiles on icy moons.
The results of our simulations are coherent with existing spectroscopic studies of Europa, both methods showing a primarily amorphous phase at the surface, followed by a crystalline phase after the first millimeter depth. Our method provides quantitative insights into how various parameters found on Europa can influence the subsurface crystallinity profiles.

Interpolating upon our simulations, we have generated a crystallinity map of Europa showing, within the top millimeter, highly crystalline ice near the equator, amorphous ice at the poles, and a mix of the two at mid-latitudes.
Regions/depths with balanced competition between crystallization and amorphization rates are of high interest due to their periodic fluctuations in crystalline fraction.
Our interpolated map reveals periodic variations, with seasonal amplitudes reaching up to $35\%$ of crystalline fraction.  These variations could be detected through spectroscopy, and we propose a plan to observe them in forthcoming missions.
\end{abstract}



\section{Introduction}

The surface of icy moons consists of water ice that can exist in either a crystalline or amorphous phase. For the typical temperature range of the Galilean moons, around $\SI{100}{K}$, water ice tends to form in its amorphous phase when it condenses from vapor \citep{Kouchi1994}.

Amorphous ices are metastable, meaning that there exists a crystalline phase with lower energy. 
The process that slowly transforms amorphous ice into its crystalline forms (cubic or hexagonal) is called thermal crystallization. The rate of crystallization is extremely sensitive to temperature, taking for example, 5 minutes at $\SI{150}{K}$ but up to one billion years at $\SI{80}{K}$ \citep{Schmitt1989}. Thus, depending on the conditions, the surface of icy moons may have crystallized or remained locked in an amorphous state.
Acting in the opposite direction, the constant bombardment of energetic particles on the surfaces increases the disorder of ice crystals, leading to amorphization \citep{Cooper2001, Baragiola2003, Berdis2020}. However, the exact mechanisms behind this radiation-induced amorphization remain unclear.

The complex balance between thermal crystallization and radiation-induced amorphization, may lead to various crystallinity on Europa depending on the dominant process. 
To determine the current state of Europa's ice, space-observation-based studies have brought valuable insights on the surface crystallinity.
\citet{Hansen2004} compared Galileo Near-Infrared Mapping Spectrometer (NIMS) ice-rich spectra of Europa's surface with models of crystalline and amorphous ice. 
On the one hand, the Fresnel surface reflection of the $\SI{3.1}{\micro m}$ band indicates that the surface's spectrum are best represented by amorphous ice.
On the other hand, the analysis of the $\SI{1.65}{\micro m}$ band shows that at depths close to $\SI{1}{\mm}$, water ice on Europa is found to be predominantly crystalline.
Complementary to this study, \citet{Ligier2016} used VLT observations to produce separate maps of the abundance of amorphous and crystalline ice, based on the $\SI{1.65}{\micro m}$ band. They did not directly provide a crystallinity map, notably because the non-ice chemicals on the trailing hemisphere complicate interpretations. Globally, they found crystalline ice to be about twice as abundant as amorphous ice. 
Ultimately, both studies are in agreement and suggest that there exists a positive vertical crystallinity gradient near the surface.
This hypothesis was recently confirmed by \citet{Cartwright2025} using the James Webb Space Telescope (JWST) NIRSpec spectrograph (1.48-$\SI{5.35}{\micro\meter}$), which identified vertically stratified crystallinity in the regolith of Europa.

Nonetheless, the presence of a band near $\SI{1.65}{\micro\meter}$ does not exclusively indicate a fully crystalline surface, as it can also appear in predominantly amorphous compositions: \citet{Mastrapa2008} demonstrated that an ice mixture made of 80\% amorphous ice can make a sample’s spectrum appear nearly fully crystalline. This highlights the limitations of linear mixture spectroscopy models for estimating accurate compositions \citep{CruzMermy2023}.

Numerical work can also be very valuable for simulating the intricate competition that leads to Europa's crystallinity.
\citet{Berdis2020} established the first modeling of the interaction between ion radiations and thermal crystallization, leading to an initial estimate of Europa's surface crystallinity. Their approach assumed a fixed average temperature for Europa's surface. 
Given the strong temperature dependency of crystallization kinetics \citep{Kouchi1994}, it would be important to consider daily temperature variations, as even brief temperature peaks within the day can significantly impact crystallization over long timescales. 
Additionally, the impact of electrons \citep{Heide1984, Lepault1983, Dubochet1988} and UV photons \citep{Kouchi1990, Leto2003, Leto2005} remains unexplored specifically on icy moons despite their significant flux \citep{Cooper2001}.

As a result, modeling studies have not yet been able to explain the vertical crystallinity gradient observed by \citet{Hansen2004} and later confirmed by \citet{Cartwright2025}, leaving several questions unanswered: 
How does crystallinity of the near surface vary for different locations on Europa? 
What are the typical depths of variations of the crystalline fraction?
The solar flux, heating the surface, varies periodically which influences the intensity of thermal crystallization. Could there be observable periodic changes in the surface crystallinity of Europa?

To accurately model ice crystallinity, these two competing processes must be 1) computed simultaneously 2) coupled to a thermal solver, given their temperature-dependency. 
To do so, they are integrated into our multiphysics surface model (MSM) \textit{LunaIcy} \citep{Mergny2024i} that incorporates a precise orbit description of Europa over long timescales, along with a fast and stable thermal solver, \textit{MultIHeaTS} \citep{Mergny2024h}, and additional physics describing the evolution of the thermal properties of the ice.

In the first section, we present the derivation of crystallization timescales and uniform irradiation dose rates for ions, electrons, and UV radiation, and their integration into the MSM. We then present the results of simulations conducted over $\SI{100 000}{years}$, covering the range of possible parameter configurations on Europa. This allows us to create an expected crystallinity map of Europa and study the temporal and depth-dependent variations in crystallinity profiles under various conditions.

\section{Methods}
\subsection{Thermal Crystallization of Ice} \label{sec:thermal_crys}

The Johnson–Mehl–Avrami–Kolmogorov equation, also known as Avrami equation, describes the kinetics of crystallization \citep{Avrami1939, Avrami1940, Avrami1941}.
According the Avrami equation, the crystalline fraction, defined as the volume fraction of crystalline ice, is given after a relaxation time $t$ by
\begin{equation}
\label{eq:avrami}
    \theta(t)=1-\exp \left( - \left( \frac{t}{\tau} \right)^n \right)
\end{equation}
where $n$ is the Avrami constant, an integer depending on the crystallization conditions, and $\tau$ is the characteristic crystallization time. 
There is no consensus in the literature regarding the exact choice of  Avrami exponent $n$ for ice crystallization on planetary surfaces.
While, for example, \citet{Kouchi1994} solves the differential equation with an exponent $n=1$, \citet{Steckloff2023} chose to use an exponent $n=4$.
In fact, the selection of this exponent depends on the conditions under which our ice crystallizes: \citet{Rao1980} compiled a table encompassing all experimentally obtained values of $n$ found in the literature. However, the characteristic time for crystallization does not depend on this Avrami exponent.

The crystallization timescale has been expressed by \citet{Kouchi1994} based on work from \citep{Seki1981}, as
\begin{equation}
    \label{eq:cristallization_time}
    \tau=\left(\frac{1}{2 \pi \alpha_{\mathrm{g}}}\right)^{1/4}\left(\frac{k_{\mathrm{B}} T}{s}\right)^{1/8} \frac{\Omega^{2/3}}{D_0} \exp \left({\frac{1}{k_{\mathrm{B}} T}}\left[E_a+\frac{4 \pi \sigma^3}{3 L_{\mathrm{c}}^2}\left(\frac{T_m}{T_m-T}\right)^2\right] \right)
\end{equation}
where $T$ is the temperature, $k_{\mathrm{B}}$ the Boltzmann constant,  and the other parameters are taken from \citet{Kouchi1994} and \citet{Schmitt1989} : $\alpha_{\mathrm{g}}=2$ is a geometric factor dependent on the growth type,  $\Omega = \SI{3.25e-29}{\cubic\meter}$ is the effective volume of a water molecule, $D_0 = \SI{6.1e-7}{\square\meter\per\second}$ is the ice self-diffusion coefficient, $E_{a} =\SI{7635e-23}{\joule}$ is the activation energy of self-diffusion, $L_{\mathrm{c}}=\SI{2.6e-21}{\joule}$ is the crystallization enthalpy per molecule, $T_m = \SI{273}{\kelvin}$ is the ice-water melting temperature, $\gamma = \SI{70e-3}{\newton\per\meter}$ is the surface tension of water and $s=\frac{2}{3} \Omega \gamma$.

For reference, the crystallization timescale is close to one billion years at $\SI{80}{K}$, around a thousand years at $\SI{100}{K}$ and less than one Europa's day at $\SI{130}{K}$. These key timescales \citep{Schmitt1989} are within the temperature range found on Europa, highlighting the necessity to accurately estimate the temperature variations.

While the Avrami Equation \eqref{eq:avrami} can be used to compute the evolution of the crystalline fraction under a constant temperature $T$, under more realistic conditions, the ice temperature varies over time, especially due to diurnal oscillations in solar flux. In such situations, it is no longer possible to use the analytical expression derived by Avrami.
To address this, \citet{Kouchi1994} suggested to approximate the crystalline fraction expression from the maximum temperature. This method assumes that only the maximum temperature affects crystallization, disregarding the entire temperature history, which significantly contributes to crystallization, particularly when temperatures are lower but close to the maximum temperature. Another drawback is the necessity of knowing in advance the maximum temperature before initiating calculations.

To improve upon this, inspired by \citet{Steckloff2023}, we propose here to discretize the differential equation that lead to Equation \eqref{eq:avrami} under the exponent $n=1$
\begin{equation} 
\label{eq:cryst_disc}
\theta (t + \Delta t) = \theta(t) + \dfrac{1 - \theta(t)}{\tau(T)} {\Delta t},
\end{equation}
where $\Delta t$ is the timestep, chosen to be greatly inferior to the crystallization timescale. Thanks to such formulation, the crystallization rate can be computed for each iteration coupled to a thermal solver, taking into account the complete temperature history.

\subsection{Radiation-Induced Amorphization}
\label{sec:amorphiz}

\subsubsection{Amorphization evolution}
The surfaces of icy moons are continuously bombarded by a range of charged ions, electrons and photons that breaks its crystalline structure. Following experimental results \citep{Moore1992, Strazzulla1992, Leto2003}, the amorphization of the crystalline fraction, when no crystallization occurs, is given by
\begin{equation}
    \theta(t)  =  \exp \left( - k(T) D(t) \right)
    \label{eq:amorph_ori}
\end{equation}
where $D$ is a dose, the radiation energy accumulated per molecule after some time $t$ and $k(T)$ is the amorphization factor.

There is no reason to expect the dose to remain constant over time. 
Here we also discretize the temporal variations for amorphization $D(t +\Delta t) = D(t) +  D(\Delta t)$ .
Using a timestep small enough such that with $k D(\Delta t) \ll 1$,  Equation \eqref{eq:amorph_ori} can be approximated to
\begin{equation}
    \theta(t+ \Delta t) \approx - \exp \left( - k(T) D(t) \right) (1- k(T)  D(\Delta t))
\end{equation}
leading to the expression
\begin{equation}
    \label{eq:amorph_disc}
    \theta(t +\Delta t) =  \theta(t) \left( 1  - k(T) D(t, \Delta t) \right).
\end{equation}
By combining \eqref{eq:cryst_disc} and \eqref{eq:amorph_disc}, the crystalline fraction can be obtained at each iteration, for any given temperature and dose.

\subsubsection{The amorphization factor $k(T)$}
\label{sec:amorph_factor}
Although energy deposition comes from different excitation sources, $k(T)$ shows a similar temperature dependence for ions
\citep{Fama2010, Baragiola2013}, referred to as $k_+(T)$ in this article.
Using the various experimental values for different temperatures and ions from \citet{Strazzulla1992}, we obtained an exponential fit 
\begin{equation}
    k_{+}(T) = \exp\left(-A_{k_{+}} T + B_{k_{+}} \right)
    \label{eq:k_amorph_fit}
\end{equation}
with, for ions, $A_{k+} = \SI{1.826e-2}{eV^{-1} K^{-1}}$ and $B_{k+} = -\SI{2.691e-1 }{eV^{-1}}$.

In contrast, experimental results suggest that for electrons and UV photons, ice becomes extremely resistant to irradiation above $\sim \SI{80}{K}$ \citep{Lepault1983, Dubochet1984, Heide1984, Kouchi1990, Moore1992, Leto2003, Loeffler2020}.
The amorphization factor for electrons was retrieved from the experimental results of \citet{Loeffler2020}, leading to the fit:
\begin{equation}
    \log k_{-}(T) = - A_{k_{-}} \exp\left(B_{k_{-}} T \right) + C_{k_{-}} 
    \label{eq:k_amorph_fit_elec}
\end{equation}
with the constants $A_{k-} = \SI{2.753e-1}{eV^{-1}}$, $B_{k-} = \SI{4.024e-2}{K^{-1}}$ and $C_{k-} = \SI{8.344e-1}{eV^{-1}}$.
Photons induced amorphization lack sufficient data with temperature \citep{Kouchi1990, Leto2003} but they show similar trends and resistance at high temperature as electron \citep{Fama2010, Baragiola2013}. For these reasons, we assume electrons and photons to share the same amorphization factor, referred to as $k_{-}(T)$,

\subsubsection{Charged particles radiations}

\begin{figure}[htpb]
	\centering
    \hspace*{-1.5cm}
	\includegraphics{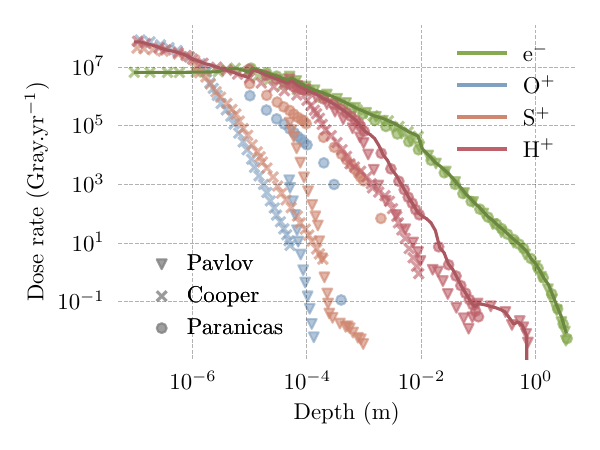}
\caption{Particle dose rates as a function of depth on Europa, derived from literature. Data points represent electrons (green), protons (red), sulfur (orange), and oxygen (blue) ions data extracted from \citet{Pavlov2018} (v), \citet{Cooper2001} (x) and \citet{Paranicas2009} (o). The combined dose rate (line) is computed by taking the maximum of interpolated values from the three datasets. Oxygen and sulfur dose rates are not included in our simulations due to high discrepancies in the literature and negligible impact compared to protons and electrons.}
    \label{fig:doses_depth}
\end{figure}

To first order, to obtain the radiation dose, various articles \citep{Cooper2001, Paranicas2001, Paranicas2009, Pavlov2018} have assumed that the deposition of energetic particles is nearly uniform over Europa's surface. For ion fluxes, the difference between hemispheres is only a few times larger on the trailing hemisphere \citep{Paranicas2002}. However, electrons show a stronger tendency: electrons with energies less than 25 MeV preferentially impact Europa’s trailing hemisphere, while those with energies greater than 25 MeV preferentially impact the leading hemisphere \citep{Cooper2001, Paranicas2009, Nordheim2018}. 

As \citet{Berdis2020} demonstrated, a one order of magnitude difference in the particle flux can alter the resulting crystallinity by approximately $10\%$.
Hence, accounting for the distribution of charged particles across Europa's surface would be highly valuable and would necessitate incorporating recent energy distributions from \citet{Addison2022} for ions and \citet{Addison2023} for electrons. Although theoretically feasible, in practice, to determine the dose at depth, this would require running simulations of the energy loss of each particle for each pair of latitude-longitude, and then MSM simulations \citep{Mergny2024i} to obtain the crystallization profiles for each latitude-longitude. This complex and time-consuming effort is beyond the scope of our first study, which focuses on establishing the competition between crystallization and amorphization.

The dose at depth is typically obtained by running Monte-Carlo simulations (e.g., SRIM \citep{Ziegler2010}, PLANETOCOSMICS \citep{Desorgher2005, Nordheim2018}), which simulate the energy loss and deflection of each charged particle due to its stopping power through the penetrated material. Such tasks have been performed over the full surface of Europa by \citet{Cooper2001}, \citet{Paranicas2009}, and \citet{Pavlov2018} for electrons, oxygen, sulfur, and protons. The extracted dose rates from these studies have been converted to $\text{Gy}\cdot\text{yr}^{-1}$ ($\SI{1}{Gy} = \SI{1}{J.kg^{-1}}$) and are plotted in Figure~\ref{fig:doses_depth}.

While the three studies do not cover the exact same range of depths, there is a very good agreement for the dose rates of electrons and protons. However, the dose rates for oxygen and sulfur differ by more than three orders of magnitude between the studies, leading us to exclude them from our analysis. This exclusion has minimal impact on our results, as even the highest estimates for oxygen and sulfur are much lower than the proton dose. The helium dose rate was not computed in any of these studies as the estimated flux of helium ions at Europa's surface is negligible compared to that of protons \citep{Mauk2004}.

To obtain the dose rate at all depths covered in our simulations, we interpolated values from the three datasets (Pavlov, Cooper, Paranicas). Where multiple values existed, we selected the maximum dose rate from one of the studies. For depths beyond the available data, the dose was set to zero. This results in the dose rate profiles for electrons and protons as shown by the solid lines in Figure~\ref{fig:doses_depth}. These dose rates, once multiplied by the timestep $\Delta t$, serve as input to compute the amorphization rate through Equation~\ref{eq:amorph_disc}.

\subsubsection{UV radiations}

As of this writing, no investigation has been conducted on the effect of UV radiation on the crystalline structure of icy moons. However, \citet{Kouchi1990} and later \citet{Leto2003} have experimentally observed the amorphization of water ice by Lyman-$\alpha$ photons ($\lambda \sim \SI{122}{nm}$).

To derive the UV-induced dose at depth,  we first define the emitted solar flux at the surface of the Sun between two wavelengths 
\begin{equation}
    F_{\odot_{\lambda_{1,2}}} (\lambda_{1}, \lambda_{2}) = \int_{\lambda_{1}}^{\lambda_{2}} B_{\lambda}(\lambda, T_{\odot}) \mathrm{d} \lambda
\end{equation}
where $B_{\lambda}$ is the spectral radiance in $\SI{}{W.m^{-3}}$  given by Planck' law
\begin{equation}
    B_{\lambda}(\lambda, T_{\odot}) = \dfrac{2 \pi h c^2}{\lambda^5} \dfrac{1}{\exp \left( \dfrac{h c}{ \lambda k_B T_{\odot}}\right) -1}.
\end{equation}
The solar flux received at the target body's surface, is then given by
\begin{equation}
    F(0, \lambda, t) =  \dfrac{R_{\odot}^2}{d(t)^2} (1-A_{\lambda})  \cos{\theta_i(t)}F_{\odot_{\lambda_{1,2}}} 
\end{equation}
where $d(t)$ is the distance to the sun, $\theta_i(t)$ the solar incidence angle and $A_{\lambda}$ is the albedo for each specific wavelength.
Europa's surface albedo in the UV spectrum was obtained from \citet{Becker2018} (Figure 8 of their article) using measurements from the International Ultraviolet Explorer, HST, and Galileo data \citep{Hendrix2005} and then interpolated over the wavelength range.

\citet{Leto2003} only demonstrated that UV-induced amorphization occurs for Lyman-alpha photons (vacuum-UV, VUV photons), which is consistent with the VUV absorption cross-section of water ice \citep{CruzDiaz2014}. Considering the optical constants from \citet{Warren2008} for wavelengths longer than approximately $\SI{160}{nm}$, the UV absorption cross-section for water ice becomes very low. Therefore, in our simulations, we consider the effective wavelength range for inducing amorphization to be $120-\SI{160}{nm}$.

It remains unclear whether other types of photons (visible, infrared, etc.) could induce amorphization, and if not, why this would be the case. Also, the amorphization rate is likely wavelength-dependent. These knowledge gaps highlight that the amorphization process is still poorly understood and would require further experimental investigation for more accurate modeling. 

Using Beer-Lambert law, the photon flux received at depth  $x$  is \citep{Lee2013}
\begin{equation}
    F(x, t) = \dfrac{R_{\odot}^2}{d(t)^2}  \cos{\theta_i(t)}  \int_{\lambda_{1}}^{\lambda_{2}} (1-A_{\lambda}) B_{\lambda} \exp \left(-\dfrac{\alpha_{\lambda}' x }{ \cos{\theta_i}(t)} \right)
\end{equation}
where $\alpha_{\lambda}'$ is the spectral absorption coefficient of porous water ice
\begin{equation}
    \alpha_{\lambda}' = \dfrac{\rho(\phi)}{\rho_0} \alpha_{\lambda} = (1 - \phi) \alpha_{\lambda}
\end{equation}
obtained from the reference absorption coefficient of pure water ice 
\begin{equation}
    \alpha_{\lambda} = \dfrac{4 \pi \Im(n(\lambda))}{\lambda}
\end{equation}
where $n(\lambda)$ is the complex index of refraction of ice Ih \citep{Warren2008}.\\

The dose can then be obtained by computing the quantity of energy $E_{r}$ absorbed by water molecules $N$
\begin{equation}
    D(x) =  \dfrac{E_{r}}{N}.
    \label{eq:def_dose}
\end{equation}
In a box of thickness $h$ and cross-section $S$, the energy received by the radiation flux $F$ after a time $\Delta t$ is given by
\begin{equation}
    E_{r} =   S \left( F(x) - F(x+h) \right) \Delta t   
    \label{eq:nrj_received}
\end{equation}
where the radiation flux is considered constant in time between $t$ and $t+\Delta t$.
The considered box, has a number of molecules equals to $N=  h \times S \times n_{\mathrm{H_2O}}$, where $n_{\mathrm{H_2O}}$ is the number density of H$_2$O molecules in the ice given by
\begin{equation}
   n_{\mathrm{H_2O}}(\phi)=\rho_0 (1-\phi)  \dfrac{ \mathcal{N}_a}{M}
   \label{eq:mol_per_vol}
\end{equation}
with $\phi$ the porosity, $\rho_0$ the density of bulk ice, $M$ the molar mass of water and $\mathcal{N}_a$ the Avogadro's constant. 
By taking the limit when the thickness of the box $h$ tends to zero, the expression of the dose absorbed by molecules in an infinitesimal layer at depth $x$ becomes
\begin{equation}
D(x, \Delta t) = - \dfrac{\partial F(x)}{\partial x}  \dfrac{\Delta t}{n_{\mathrm{H_2O}}}
\label{eq:dose_deriv_flux}
\end{equation}
As a result, the UV-induced dose at depth is obtained from Equation \eqref{eq:dose_deriv_flux}, 
\begin{equation}
D_{\mathrm{UV}}(x, \Delta t) =  \dfrac{R_{\odot}^2}{d(t)^2}  \dfrac{\Delta t} {n_{\mathrm{H_2O}}(\phi)} \int_{\lambda_\mathrm{1}}^{\lambda_{\mathrm{2}}}  \alpha_{\lambda}'    (1-A_{\lambda})  B_{\lambda}   \exp \left(-\dfrac{\alpha_{\lambda}' x }{  \cos{\theta_i(t)} } \right)\mathrm{d} \lambda 
\end{equation}
which is coherent with the expression found in \citet{Cook2007}.

\subsection{Coupling with \textit{LunaIcy}}

The thermal crystallization process is highly dependent on the temperature. 
As both crystallization and amorphization processes occur competitively on the surface and subsurface of icy moons, it is necessary to compute them simultaneously.
Therefore, to accurately simulate the crystalline evolution, we have integrated the crystallization and amorphization modules into \textit{LunaIcy} \citep{Mergny2024i}.
The model consists of a uni-dimensional bloc of ice made of grains, with a certain porosity and thermal properties for each depth. 
In this numerical model, we use an irregular spatial grid consisting of $n_x$ points, which we iterated for a total of $n_t$ iterations. 

By calculating the orbit of the target body, we determine the solar flux heating the ice surface. Then, our thermal solver \textit{MultIHeaTS} computes the heat transfer throughout the material's depth for each timestep. 
At each depth, the crystalline module assesses the crystallinity changes corresponding to the current temperature and radiation flux. This iterative process is repeated throughout the simulation. 
A complete description of \textit{LunaIcy} is presented in detail in \citep{Mergny2024h, Mergny2024i}.


The \textit{LunaIcy} model is run over a range of parameters to simulate the evolution of Europa's icy surface crystallinity under various conditions for $\SI{100 000}{years}$.
While temperature variations occur within the diurnal period, crystallization and amorphization processes can occur over large timescales under Europa's conditions.
To capture temperature variations, the simulations require a precise timestep $\Delta t = p_e / n_{\mathrm{spd}}$, where $p_e = \SI{3.55}{days}$ is Europa's orbital period and here $n_{\mathrm{spd}} = 60$.

The determination of the initial ice structure is difficult due to the lack of information on Europa. 
As a reference scenario, we choose to impose a constant thermal inertia of $\SI{95}{m^2 K s^{1/2}}$ everywhere on the moon, following results from \citet{Trumbo2018} and \citet{Rathbun2010}.
Also, we assume that ice is formed in its crystalline phase on Europa's surface, leading to an initial crystalline fraction of $\theta(t=0) = 1$, a scenario that occurs, for example, when liquid water from cryovolcanism freezes at the surface. 
While an initial condition of $\theta(t=0) = 0$ could also be used, the cold regions of Europa would not have sufficient time to crystallize during the simulation, preventing the system from reaching equilibrium. 
A summary of the model parameters and initial conditions can be found in Table~\ref{tab:params}.

Please note that some aspects of the coupling have been neglected. Studies, such as \citet{Gudipati2013}, have found that the thermal conductivity of amorphous ice is lower than that of crystalline ice, although it is unclear whether this is due to differences in porosity. Additionally, \citet{Raut2008} showed that nanoporous ice compacts due to ionizing radiation in laboratory results and \citet{Palumbo2010} indicated that both ion and UV irradiation could slightly alter the porosity of amorphous water ice. Finally, \citet{Steckloff2023} also mentioned that the transition from amorphous to crystalline ice is exothermic, which should be included in the heat budget. Although we acknowledge these differences between crystalline and amorphous ice, they do lead to important changes in the thermal properties and will be addressed in future refinements of the model.

\begin{table}[htbp]
    \centering
    \begin{tabular}{ll}
    \toprule
    Parameters & Value  \\ 
    \midrule
    Simulations time  & $\SI{100 000}{years}$ \\
    Timestep $\mathrm{dt}$ & \SI{85}{min}\\
    Grid points $n_x$ & 100  \\
    Initial crystallinity $\theta(t=0)$ & 1 \\
    Thermal inertia  &  $\SI{95}{m^2 K s^{1/2}}$  \\
    Porosity $\phi$ & 0.5  \\
    Longitude $\psi$ & 180 \\
    Latitudes $\lambda$ & (0 - 80) \\
    Bond albedos $A$ & (0.3 - 0.8)   \\
    \bottomrule
    \end{tabular}
\caption{Main physical and numerical parameters used, along with initial conditions.}
\label{tab:params}
\end{table}

\section{Results}
\label{sec:results}

\subsection{Radiation Dose Profiles}

Although the dose rates shown in Figure~\ref{fig:doses_depth} are necessary inputs to our model,  they do not provide a clear representation of the amorphization they induce. As shown by Equation~\ref{eq:amorph_ori}, the crystallinity $\theta$ depends not only on the dose but also on the amorphization factor $k(T)$, which can vary by orders of magnitude across the temperature range found on Europa (see Section \ref{sec:amorph_factor}).

For visualization purposes, it is more insightful to estimate for each radiation source the dosage time $t_{D_{\mathrm{ef}}}$ required to reach an efficient amorphization dose $D_{\mathrm{ef}}$, defined by $k(T) D_{\mathrm{ef}} = 1$. The dosage time $t_{D_{\mathrm{ef}}}$ represents the time it takes for the crystalline fraction to decrease by a factor of $1/e$.

\begin{figure}[htpb]
	\centering
    \hspace*{1.0cm}
	\includegraphics{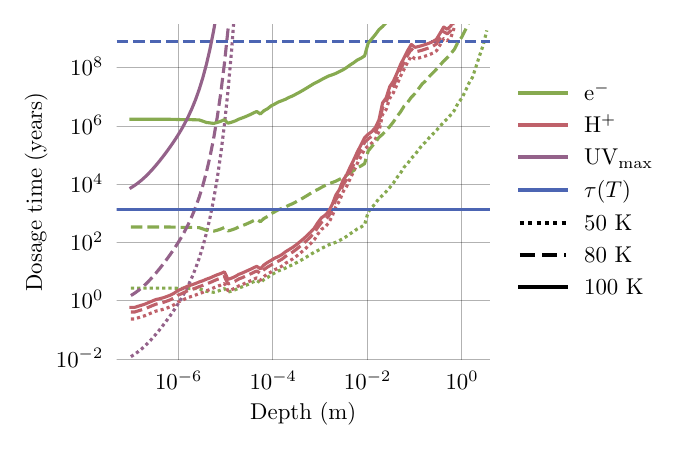}
\caption{Efficient dosage time for electrons (green), protons (red), and maximum daily UV (purple) radiations as a function of depth on Europa, for three different temperatures: $\SI{50}{K}$ (solid line), $\SI{80}{K}$ (dashed line), and $\SI{100}{K}$ (dotted line). The dosage time is defined such that $k(T) D_{\mathrm{ef}}( t_{D_{\mathrm{ef}}}) = 1$, representing the time it takes for the crystalline fraction to decrease by a factor of $1/e$. For reference, the crystallization timescale is shown at $\SI{80}{K}$ and $\SI{100}{K}$ (with $\SI{50}{K}$ being too cold for any relevant crystallization to occur).}
    \label{fig:efficient_dose}
\end{figure}

Since $k(T)$ varies with temperature and radiation sources, we have plotted in Figure \ref{fig:efficient_dose} the dosage time versus depth for electrons, protons, and UV radiations at three different temperatures: $\SI{50}{K}$ (solid line), $\SI{80}{K}$ (dashed line), and $\SI{100}{K}$ (dotted line). These temperatures were selected because $\SI{50}{K}$ is close to the minimum temperature on Europa, $\SI{80}{K}$ is the temperature above which the amorphization efficiency of electrons and UVs becomes negligible, and $\SI{100}{K}$ is the mean temperature of Europa \citep{Ashkenazy2019}. For each temperature, the crystallization timescale, computed from Equation~\ref{eq:cristallization_time}, is shown for comparison, except for $\SI{50}{K}$, as $\tau(\SI{50}{K})$ exceeds the age of the universe.

From Figure~\ref{fig:efficient_dose}, we observe that although the electron flux at the surface of Europa is the highest, it leads to efficient amorphization only at low temperatures, such as $\SI{50}{K}$, e.g., at high latitudes during the night.

Over all temperature ranges and depths, proton dose is the most dominant form of particle radiation. On a timescale of up to 100,000 years, which we chose for our simulations, if no crystallization were occurring, proton radiation would lead to efficient amorphization up to the cm depth.  However, since thermal crystallization occurs simultaneously, to correctly estimate  the actual state of crystallinity, coupled simulations are required in order to assess the outcome of these competing processes.

UV-induced dosage times are also shown for a reference case at the equator, using Europa's mean distance to the Sun, $\SI{5.20}{AU}$, and considering the highest daily value. Due to the high absorption of Europa's surface in the UV spectrum \citep{Becker2018}, UV radiation is absorbed in the first superficial layers, resulting in high doses at very shallow depths ($<\SI{1}{\micro m}$). However, the behavior of the UV's amorphization factor with temperature remains unclear, and further experimental work is needed to better constrain the amorphization rate induced by UV radiations.

Although we have assumed a uniform distribution of particles, in reality, different locations on the surface receive particles of varying energies \citep{Addison2023}. Since penetration depth is energy-dependent \citep{Teolis2017}, this would result in different dosage times at different locations.

\subsection{Parameter Exploration}
\begin{figure}[htpb]
	\centering
    \hspace*{1.cm}
	\includegraphics[width=\textwidth]{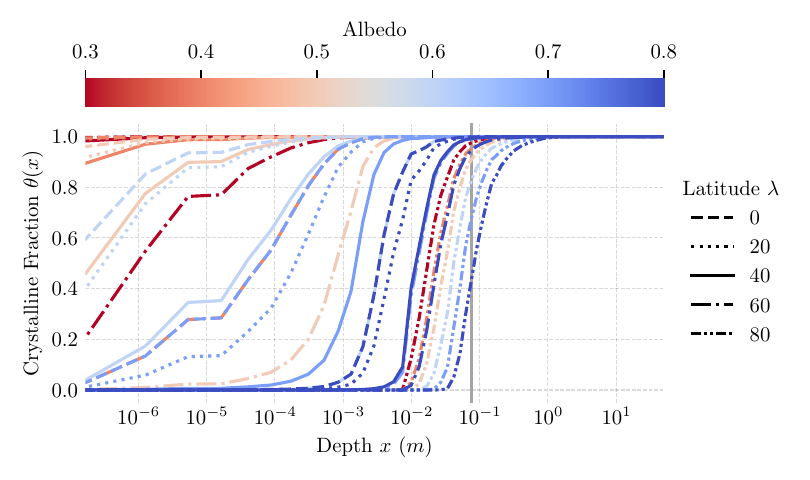}
 \caption{Crystalline fraction profiles after $\SI{100 000}{years}$ for the set of simulations of Europa. We vary the albedo from 0.3 to 0.8 with a step of 0.1, while latitude ranges from 0 to 80 degrees with a step of 20 degrees, leading to  30 different simulations. The diurnal thermal skin depth is indicated by the gray line at $\sim \SI{7}{cm}$. }
    \label{fig:all_thetas}
\end{figure}

Thanks to the numerical model \textit{LunaIcy}, we can simulate all conditions affecting crystallinity on Europa by conducting a parameter exploration.
While the charged particles flux is considered uniform in this study, the UV dose changes with latitude. Moreover, thermal crystallization is temperature dependent and hence will depend on the solar flux, surface albedo and thermal properties.
To investigate this parameter dependency, a set of simulations is run with varying albedo values from 0.3 to 0.8 with a step of 0.1, covering the range found on Europa \citep{Rathbun2010}. 
Simultaneously, solar flux is latitude dependent, so latitude values are varied from 0 to 80 degrees with a step of 20 degrees, as negative latitudes do not need to be computed due to the symmetry of the solar flux on Europa.
The ice thermal properties are fixed, with the reference thermal inertia $\SI{95}{m^2 K s^{1/2}}$ \citep{Trumbo2018}.
This parameter exploration results in 30 different simulations, representing various thermal configurations found on Europa.
The final crystalline fraction profiles after $\SI{100 000}{years}$ for the 30 albedo/latitude pairs are shown in Figure \ref{fig:all_thetas}.

As expected, the coldest regions, at high latitudes and high albedo, maintain completely amorphous ice.
Also anticipated, hottest regions, at low latitudes and very low albedo, present an almost completely crystalline profile after $\SI{100 000}{years}$. 
Thermal crystallization is simply too efficient at these temperatures, and simultaneously, the amorphization efficiency of particles and UVs is significantly decreased. 

Regions of particular interest are where a balanced competition between thermal crystallization and amorphization can happen.
Such scenario is often seen at mid latitudes, for example, for the simulation $\lambda=40$\textdegree, $A=0.6$ (see Figure \ref{fig:all_thetas}, lightest blue full line).
The radiation dose predominates at shallow depths ( $<\SI{1}{\micro m}$) and then gradually decreases with depth (see Figure \ref{fig:doses_depth}) to allow for a balanced competition with thermal crystallization. 
Around the $\sim \SI{1}{mm}$ depth, the dose has significantly decreased, allowing thermal crystallization to dominate, leading to the crystallization peak at the $\sim \SI{1}{mm}$ depth (see Figure \ref{fig:all_thetas}).
This result is particularly interesting because it is in agreement with the spectroscopy-based model \citep{Hansen2004}  that found ice at the $\SI{1}{mm}$ depth on Galilean moons to be predominantly crystalline.

Beyond the diurnal thermal skin depth ($> \SI{7}{cm}$), daily solar variations are greatly attenuated, but simultaneously most of the radiation doses have already been absorbed by shallower layers. As a consequence, the ice at these depths remains crystalline.

Note that, due to computational constraints, simulations are run for $\SI{100 000}{years}$; however, within this timeframe, some of the hottest regions may not have sufficient time to amorphize. 
In fact, for low dose rates, the amorphization timescale (see Figure \ref{fig:efficient_dose}) can exceed $\SI{100 000}{years}$. In most cases, these low doses are rapidly overshadowed by faster crystallization timescales (see Figure \ref{fig:efficient_dose}). 
In the few cases where temperatures are low enough (\textit{i.e.}, at the poles), this limitation means that we locally underestimate the amorphization. However, this underestimation occurs only at depths not probed by current spectroscopy methods for resolving crystallinity (e.g., deeper than the first millimeter).

Overall our numerical results are in great agreement with spectroscopic observations of a vertical crystallinity gradient on Europa's surface \citep{Hansen2004, Ligier2016}.
Our model shows that high radiation doses are absorbed at the very near surface, resulting in a high fraction of amorphous ice within the first micrometer (except in the hottest regions). The same interpretation was derived from spectroscopic observations using the Fresnel surface reflection band at $\SI{3.1}{\micro m}$, which \citet{Hansen2004} interpreted as evidence of a predominantly amorphous ice surface.
Moreover, our results are also consistent with spectroscopic observations of the $\SI{1.65}{\micro m}$ band, which probes the first millimeter depth and supports the use of crystalline ice to fit Europa's spectra at these depths \citep{Hansen2004, Ligier2016}.

The concept of a vertical crystallinity gradient was proposed by \citet{Hansen2004} and \citet{Ligier2016} to explain the differences in crystallinity observed between the $\SI{1.65}{\micro m}$ and $\SI{3.1}{\micro m}$ bands on Europa's spectra. Through our numerical modeling, we have demonstrated how the competition between processes can lead to such crystallinity profiles. This highlights an important consideration when representing Europa's surface: each depth will lead to a different crystallinity map.

\subsection{Crystallinity Map}

\paragraph{Mapping Crystallinity within the First Millimeter}

\begin{figure}[htpb]
	\centering
	\includegraphics[width=\textwidth]{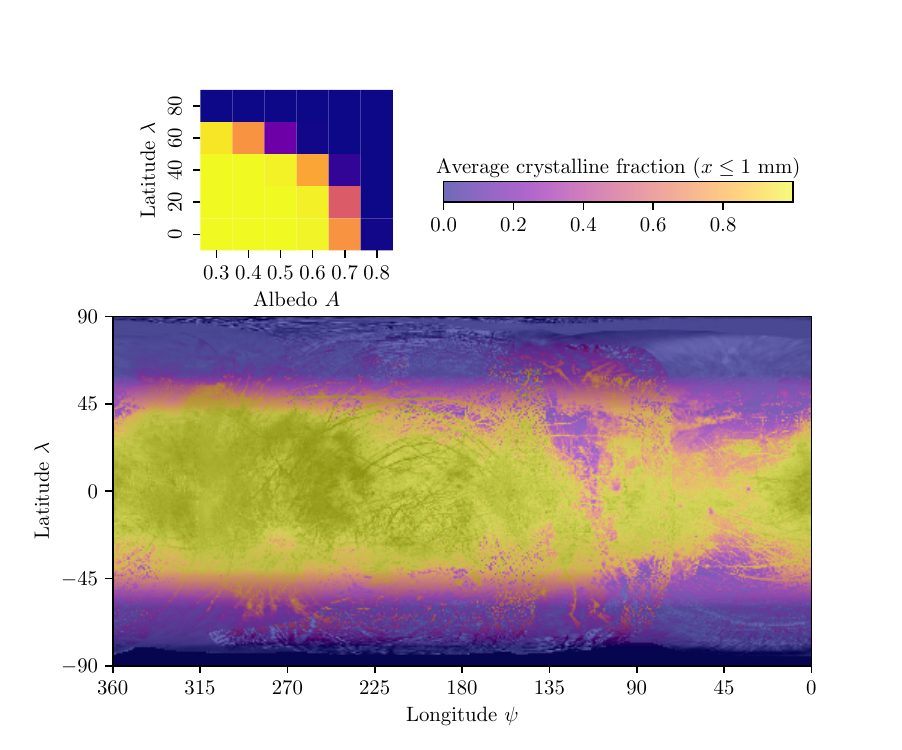}
 \caption{(\textit{Top Left}) Average crystalline fraction heatmap for depths $<\SI{1}{mm}$ computed on Europa for a uniform flux of particles as function of albedo and latitude. (\textit{Bottom}) Interpolation of the averaged crystalline fraction heatmap to the albedo map of Europa. }
    \label{fig:cry_lat_alb}
\end{figure}

Various observable quantities can be derived from the crystallization profiles shown in Figure \ref{fig:all_thetas}.
Interpretation of the infrared spectra from the Near Infrared Mapping Spectrometer on Galileo \citep{Hansen2004} led to estimations of the crystalline structure of Europa, with a critical depth $\sim \SI{1}{mm}$ where a transition from amorphous to crystalline ice has occurred.
To compare with spectroscopy measurements, we derive the average crystalline fraction after 100 kyr on the first $d=\SI{1}{mm}$ depth:
\begin{equation}
    \langle  \theta \rangle (d)  = \dfrac{1}{d} \int_0^d \theta(x) dx.
\end{equation}
The average crystalline fraction within the first millimeter, denoted as $\langle \theta \rangle_{\mathrm{mm}}$, is displayed for the albedo/latitude parameter exploration in Figure \ref{fig:cry_lat_alb} (\textit{Top Left}).
Consistent with Figure \ref{fig:all_thetas}, crystallinity decreases from fully crystalline to entirely amorphous with increasing latitude and albedo.

Then, the first-millimeter averaged crystalline fraction $\langle \theta \rangle_{\mathrm{mm}}$ computed for discrete values (Figure \ref{fig:cry_lat_alb}, \textit{Top Left}), is interpolated over a continuous range of albedos $A \in [0.3, 0.8]$ and latitudes $\lambda \in [-\ang{90}, \ang{90}]$.
This interpolation allows for the computation for all albedo/latitude pair values found on Europa's reconstructed albedo map, as illustrated in Figure \ref{fig:cry_lat_alb} (\textit{Bottom}).
The fitted albedo map of Europa was previously generated \citep{Mergny2025} by calculating the Bond albedo across 20 regions of interest on Europa \citep{Belgacem2020}, combined with visible data from the USGS global mosaic \citep{Becker2010}.
Note that a data gap exists in the Europa mosaic at the south pole, with no coverage below latitude -83\textdegree, and only low-resolution data cover the north pole and many high latitudes in both the north and the south.

\paragraph{Trend, Regions of Interest and Spectroscopy}

Results show that, at the first millimeter depth, near equatorial regions $\lambda \in [-\ang{30}, \ang{30}]$ of the trailing hemisphere are fully crystalline, whereas the leading hemisphere displays more variability due to higher albedo.
Irrespective of longitude, high-latitude regions $|\lambda|> \ang{60}$ are characterized by entirely amorphous ice, attributed to the slow thermal crystallization at these temperatures.
Regions displaying a mixed amorphous/crystalline ratio are particularly noticeable at mid latitudes $\lambda \sim \ang{50}$, experiencing a balance between thermal crystallization and irradiation-induced amorphization.
Notable high albedo features, like Pwyll crater (25.2 \textdegree S 271.4 \textdegree W) have locally lower crystallinity due to the lower temperature.
The lower crystallinity values obtained at the northern low latitudes of the trailing hemisphere (10--30\textdegree N, 105\textdegree W) align very well with recent observations from JWST NIRSpec \citep{Cartwright2025}, due to the region's higher albedo and lower surface temperature.

The crystallinity map shown in Figure \ref{fig:cry_lat_alb}, is of particular interest for the analysis of existing spectroscopy measurements and the suggestion of target spots for the upcoming spatial missions.
We suggest comparing spectroscopy measurements between low latitude, low albedo regions like Dyfed Regio (10\textdegree N, 250\textdegree W) and high latitude regions like Balgatan Regio (50\textdegree S, 30\textdegree W).
Our model would be confirmed if distinct crystallinity profiles were observed, with completely amorphous ice at the poles and fully crystalline ice near the equator.

However, due to the different composition and microstructure of the ice in different regions, cautious work must be made to properly estimate the crystallinity from spectroscopy measurements.
As shown by \citet{CruzMermy2023}, various combinations of materials can explain the same spectrum and thus can lead to different abundance of crystalline/amorphous H$_2$O.
Also, as the authors have shown, a mix between amorphous and crystalline ice results in non linear optical properties, and thus one must be careful when estimating a crystalline fraction from spectroscopy. 

While \citet{Ligier2016} presented composition maps for crystalline and amorphous ice, several of their underlying assumptions make these maps difficult to interpret in terms of crystallinity. As illustrated in Figure 11 of their paper, the presence of salts on the surface systematically reduces the detected amounts of crystalline or amorphous water ice due to the decreased amount of water. As a result, the crystalline ice fraction on their map does not correlate positively with albedo values; low albedo regions, which are hotter, counterintuitively show less crystalline ice, simply because they contain less water.
To be able to compare our map with \citet{Ligier2016}, we would need their crystalline and amorphous ice compositions to be combined and normalized by the total amount of water ice, into a crystallinity map.

It would be highly unexpected to observe any significant crystallinity near the poles since crystallization is inefficient in those regions, unless our understanding of Europa's surface temperature is significantly mistaken, which seems unlikely. 
The presence of high crystallinity at the poles would necessitate two conditions to be met: first, significantly lower radiation levels or a less efficient amorphization process than anticipated, and second, frequent mechanisms such as surface recycling or deposition of ice at temperatures high enough to be renewed in its crystalline state.

If lower crystallinity is detected in the first millimeter near the equator of the trailing hemisphere, a potential explanation would be to consider the high flux of low-energy electrons ($< \SI{10}{keV}$) as a factor inducing amorphization. 
To verify this, crystallinity near the equator of the leading hemisphere could also be measured, for example, in Tara Regio (10 \textdegree S 75\textdegree W), where it is assumed that no low-energy electrons reach the surface \citep{Paranicas2009, Nordheim2017}.
If amorphous ice is also observed in that region, it would either suggest that other sources of amorphization, such as ions or UV radiation, may be more numerous or efficient than expected, or less likely, that our current estimations of Europa's surface temperatures \citep{Rathbun2010, Ashkenazy2019} and thermal properties \citep{Trumbo2018} are significantly different from reality.

\paragraph{Limitations and Outcomes}
Several assumptions needed to be made to allow such theoretical modeling. These may result in a crystallinity map that is different in reality.
For irradiation:
\begin{itemize}
    \item[-] We assumed a constant flux of charged particles, yet several studies \citep{Paranicas2009, Teolis2017, Nordheim2017} suggest significant variations in electron flux reaching icy moons surfaces across different locations, with only MeV-energy electrons able to reach the leading hemisphere, resulting in less amorphization at shallow depths there. Variations in ion flux with latitude/longitude should also be considered, but they are less pronounced and would induce minor fluctuations in crystallinity.
    \item[-] We considered a surface made of pure H$_2$O ice, whereas Europa’s surface includes non-water ice components \citep{McCord1999, Carlson2005, Ligier2016, CruzMermy2023}, which would lead to different radiation penetration depths of particles if incorporated.
\end{itemize}
For crystallization:
\begin{itemize}
  \item[-]  Due to the lack of high-resolution data in current observations of Europa, the reconstructed albedo map from \citet{Mergny2025} includes regions with very low resolution, leading to less accurate albedo estimates in those areas.
  \item[-]  The effects of diurnal eclipses on the sub-jovian hemisphere were not considered. Adapting the approach from \citet{Mergny2024h} to modify the albedo map could be applied to model the influence of eclipses on the thermal balance and thus crystallization rates.
  \item[-]  A uniform thermal inertia was assumed, while taking into account variations across the surface will result in local changes of the crystallization timescales. However, current estimates of Europa’s thermal inertia rely on assumptions of homogeneous material \citep{Rathbun2010, Trumbo2018}, while Europa’s surface is likely vertically heterogeneous \citep{Hansen1973, Mergny2024h}. More advanced thermal analysis is needed to fully account for these effects.
\end{itemize}
We anticipate that upcoming missions JUICE \citep{Grasset2013} and Europa Clipper \citep{Phillips2014} will provide better surface observations, enabling better constraints on these parameters and leading to improved modeling.

\subsection{Periodic Variations}

\begin{figure}[htpb]
	\centering
	\includegraphics[width=\textwidth]{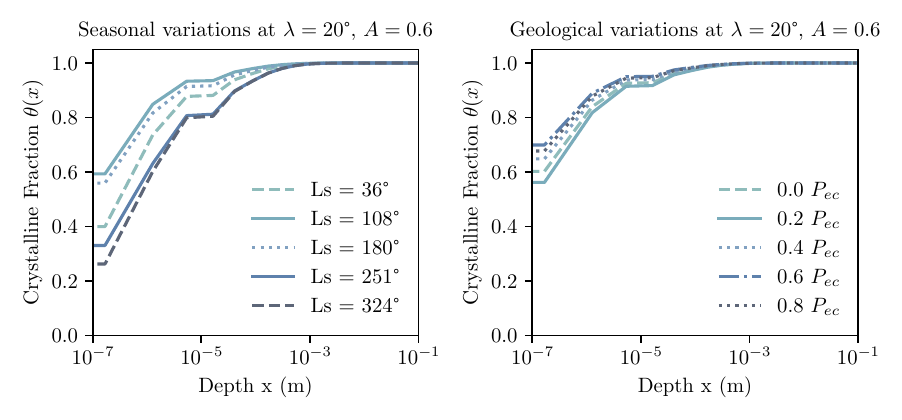}
    \caption{ (\textit{Left}) 
    Crystalline fraction profiles for various solar longitude, for a surface at latitude $\lambda = \SI{20}{\degree}$, albedo $A = 0.6$ and $Lt = \SI{0}{\degree}$. Seasonal variations are observable at greater depth and amplitude for mid-latitude regions where there exist an equilibrium between particle amorphization and crystallization.
    (\textit{Right})
    Crystalline fraction profiles of the same surface for different locations during the eccentricity oscillation period $P_{\mathrm{ec}}$ at the same local time and solar longitude $Ls = Lt = \ang{0}$. \textit{LunaIcy}'s orbital model captures Europa's eccentricity period of about $P_{\mathrm{ec}} \sim \SI{55 000}{years}$, leading to temperature fluctuations that affect the crystalline profiles.
    }
    \label{fig:var_theta_double}
\end{figure}

The solar flux heating the surface of Europa has diurnal, seasonal and geological fluctuations that lead to varying efficiency of the thermal crystallization.
An unprecedented implication of this is that thanks to the simulated competition between crystallization and radiation, we can observe periodic variations of the crystallinity profile.

\paragraph{Diurnal Variations}

To observe diurnal variations in crystallinity, much shorter dosage times would be required than those implemented in our model, which would amorphize the cold surface of the moon during the night. In contrast, during the day, if the same surface experiences temperatures above $\SI{130}{K}$, it will undergo thermal crystallization within a timescale shorter than Europa's day length, leading to an increase in surface crystallinity. Based on the dosage times shown in Figure~\ref{fig:efficient_dose}, it seems very unlikely that such diurnal variations in crystallinity can occur.

\paragraph{Seasonal Variations}
While Jupiter's tilt is only $\ang{3}$, due to its eccentricity $e = 0.048$, the surfaces of Galilean moons experience seasonal variations of the solar flux.
These variations can lead to surface temperatures variations of up to $\SI{5}{K}$ on Europa \citep{Ashkenazy2019} and lead to significant changes on the crystallization rate.
For regions where there is a delicate balance between crystallization and radiations, this can lead to important changes of the crystallinity profiles.
In Figure (\ref{fig:var_theta_double}, \textit{Left}), is shown the crystallinity profiles for various locations in Jupiter's orbit, for the same surface of Europa, compared at the same local time. 
Near the depth of $d \sim \SI{1}{\mu m}$, there are considerable variations in the crystalline fraction (up to $35 \%$)  at different solar longitudes.

Since these variations are most pronounced near the surface, they could potentially be observed through the Fresnel reflection of the surface, using the $\SI{3.1}{\micro m}$ band as a proxy for crystallinity \citep{Hansen2004, Ligier2016}.
To anticipate comparison with spectroscopy measurements, we calculate the average variations in crystallinity at the first $d=\SI{1}{\micro m}$ depth: 
\begin{equation}
\langle \Delta \theta \rangle (d) = \dfrac{1}{d} \int_0^d \max_{\mathrm{Ls}}{ \left| \Delta \theta(x) \right|} dx.
\end{equation}
This represents the average over the first micron of the maximum peak-to-peak difference between crystalline fraction profiles of different solar longitudes.
These seasonal fluctuations are illustrated in Figure (\ref{fig:cry_lat_alb}, \textit{Top Left}) as part of the albedo/latitude parameter exploration. 

\begin{figure}[htpb]
	\centering
	\includegraphics[width=\textwidth]{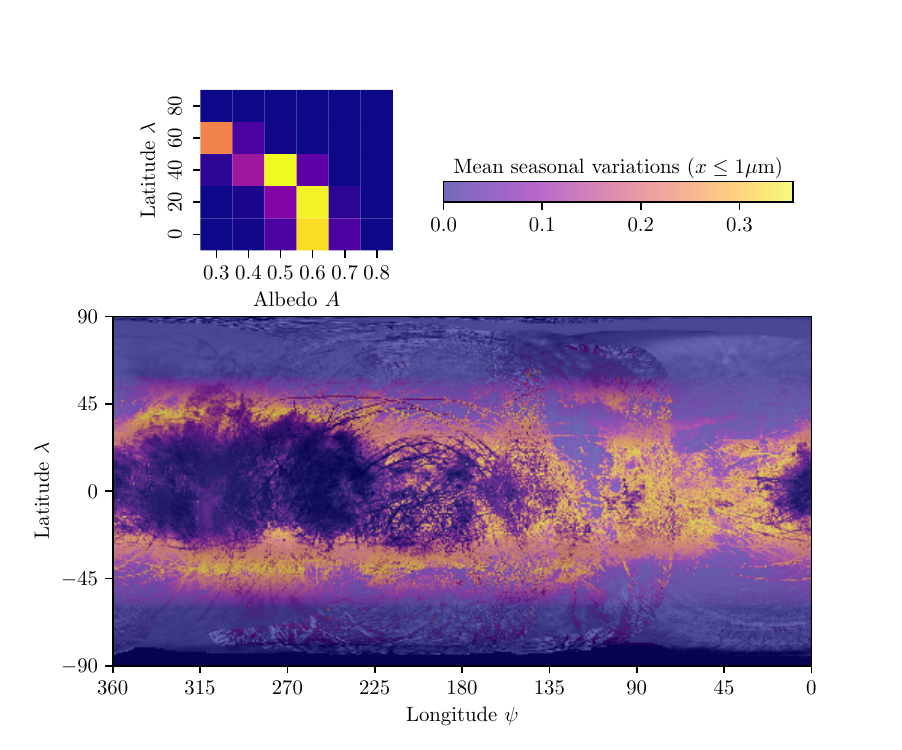}
 \caption{(\textit{Top Left}) Heatmap of the mean seasonal variations for the very near surface of Europa (x $<\SI{1}{\micro m}$) as function of albedo and latitude. (\textit{Bottom})  Interpolation of the seasonal variations heatmap to the albedo map of Europa. }
    \label{fig:season}
\end{figure}

Following the previous method, the seasonal variation heatmap is then interpolated over a continuous range of albedos $A \in [0.3, 0.8]$ and latitudes $\lambda \in [-\ang{90}, \ang{90}]$.
This allows us to generate a map of mean seasonal crystallinity variations in the first millimeter depth across Europa, as shown by Figure (\ref{fig:season}, \textit{Bottom}).
It is evident from this map that regions with pronounced seasonal fluctuations are globally situated at mid-latitudes, around $\lambda \sim \ang{40}$ and for the leading hemisphere, also at low-latitudes. 
While equatorial regions of the trailing hemisphere are dominated by high thermal crystallization locking the ice into its crystalline state, the higher-albedo leading hemisphere  experiences high seasonal variations due to a more balanced competition with radiations.
However, as discussed previously, simplifications in our model may lead to differences between the presented map and the actual seasonal variations on Europa's surface.
Given the delicate balance required for seasonal variations, uncertainties in modeling are even more particularly sensitive here.
Nonetheless, considering the fast crystallization rate of near-equator areas and the locked amorphous state at the poles, mid-latitude regions likely represent the most favorable areas for a balanced competition.

This is particularly interesting for the upcoming missions, as the Europa Clipper's expected lifespan of $\SI{4}{years}$ aligns well with the potential observability of these variations.
Given Jupiter's orbital period of $p_\mathrm{J}  = \SI{11.86}{years}$, the highest crystallinity variations occur over approximately $\SI{5.43}{years}$ (half a period).
This means that considerable variations could be detected within the Europa Clipper's operational timeframe. As shown in Figure \ref{fig:var_theta_double} (\textit{Left}), at $\lambda = \SI{20}{\degree}$ and $A = 0.6$, near the micrometer depth, the crystallinity varies from $\theta \approx 0.45$ during the northern hemisphere winter to $\theta \approx 0.8$ during the summer, indicating that variations of $\sim 35\%$ could be observed.
For example, we suggest conducting multiple observations over several years at the same local time within regions such as the Pwyll crater (25.2° S, 271.4° W) or the Tara Regio (10° S, 75° W). 
These areas are expected to show significant seasonal variations and are also of particular interest for other scientific purposes \citep{Villanueva2023, Trumbo2023}.

\paragraph{Geological Variations}
Over large timescales, some of the orbital parameters, notably the eccentricity of the Galilean system, fluctuate with periods ranging from 27,000 to 1.1 million years \citep{Laskar2008},  which is accounted in \textit{LunaIcy}'s orbital module \citep{Mergny2024h}.
The most pronounced period occurs approximately every $P_{\mathrm{ec}} \sim \SI{55,000}{years}$ and induces small variations of the solar flux.
Given that crystallization strongly depends on temperature, these fluctuations lead to noticeable changes in crystallinity profiles.
Figure (\ref{fig:var_theta_double}, \textit{ Right}) shows the crystalline fraction profiles at various locations over the eccentricity oscillation period $P_{\mathrm{ec}}$, at latitude $\lambda = \SI{20}{\degree}$, albedo $A = 0.6$, and local time $Lt = \SI{0}{\degree}$.
Variations of up to $\SI{10}{\%}$ in crystallinity are noticeable between these different locations.
While these fluctuations are not observable within our lifespan, they suggest that crystallinity profiles could provide insights on the surface's thermal history.

\section{Conclusion and Perspectives}

Many processes affecting the surfaces of icy moons, such as thermal crystallization or amorphization, are highly temperature-dependent and cannot be accurately estimated with a fixed temperature.
For these reasons, they must be coupled with a precise temperature evolution,  which is why  \textit{LunaIcy} was developed.
Competing with this process, radiation of charged particles can lead to amorphization of ice, but also from electrons and UV on whose effects on crystallinity have not been explored for Europa yet. 

To better understand which process dominates on Europa, we conducted a set of MSM simulations where both thermal crystallization and radiation-induced amorphization are integrated into \textit{LunaIcy}.
Through a parameter exploration covering relevant latitudes and albedo values, we explored various configurations of the icy moon.

Our results showed that under most conditions on Europa, the surface is primarily amorphous ice, transitioning to a crystalline phase around the millimeter depth, which aligns well with existing spectroscopy observations of Europa \citep{Hansen2004, Ligier2016, Cartwright2025}.

Interpolating these results allowed us to generate a crystallinity map for the first millimeter depth across Europa. Our current model reveals a transition from amorphous ice at the poles to fully crystalline at the equator, with a mixture of the two at mid-latitudes.
Although our model assumes a uniform distribution of charged particles, in reality electrons are mainly concentrated on the trailing hemisphere. Accounting for this asymmetry in future studies may lead to a decrease in crystallinity on the trailing hemisphere and an increase on the leading hemisphere.

Remarkably, the simulations of these competing processes have revealed periodic variations in the crystallinity profiles. 
Notably, seasonal variations have the highest amplitudes, reaching crystallinity fluctuations of up $\SI{35}{\%}$ in mid-latitude regions. 
This is significant, as upcoming missions like Europa Clipper and JUICE could potentially observe these seasonal variations during their operational lifetimes.
Additionally, smaller yet noticeable variations are expected on timescales of $\sim \SI{55 000}{years}$.

This study has focused on the coupled modeling of crystallization, amorphization, and heat variations to quantitatively estimate crystallinity profiles at various depths under a range of conditions found on Europa. This approach necessitated several simplifications, which may introduce discrepancies with reality.
We propose the observation of key regions—Dyfed Regio, Balgatan Regio, Pwyll Crater, and Tara Regio—during upcoming missions Europa Clipper and JUICE to validate or refine our model. Improved spectroscopy models will also be essential for accurately estimating crystallinity within acceptable error margins given the nonlinear nature of the process.

\section*{Statements and Declarations}
\subsection*{Funding and Competing Interests}

We acknowledge support from the ``Institut National des Sciences de l'Univers'' (INSU), the ``Centre National de la Recherche Scientifique'' (CNRS) and ``Centre National d'Etudes Spatiales'' (CNES) through the ``Programme National de Plan{\'e}tologie''. 

The authors have no competing interests to declare that are relevant to the content of this article.

\section*{Data Availability}
The authors assert that the data supporting the study's findings are included in the paper, while some of the code is accessible on the online public repository at \url{https://gitlab.dsi.universite-paris-saclay.fr/cyril.mergny/MultIHeaTS}, with the remainder available upon request.

\section*{Acknowledgments}
We acknowledge support from the ``Institut National des Sciences de l'Univers'' (INSU), the ``Centre National de la Recherche Scientifique'' (CNRS) and ``Centre National d'Etudes Spatiales'' (CNES) through the ``Programme National de Plan{\'e}tologie''. We gratefully acknowledge the support of the CNES Computing Center, which provided access to HPC resources including TREX and the GEOPS laboratory for access to the BMO supercomputer, where parts of the simulations were conducted.

\bibliography{library}
\bibliographystyle{aasjournal}

\end{document}